\documentclass[preprintnumbers,secnumarabic,amsmath,amssymb,nofootinbib,floatfix]{revtex4}
\usepackage{varioref,exscale,latexsym,amsmath,amssymb}
\usepackage{graphicx,epstopdf}
\usepackage{comment}  
\usepackage{epsfig}
\usepackage{graphicx}% Include figure files
\usepackage{dcolumn}% Align table columns on decimal point
\usepackage{bm}% bold math
\usepackage{simplewick}

\def\beq{\begin{equation}}

\def\eeq{\end{equation}}

\def\beqa{\begin{eqnarray}}

\def\eeqa{\end{eqnarray}}

\begin{document}

\title{{\bf QED trace anomaly, non-local Lagrangians and quantum Equivalence Principle violations}}

\medskip\
{\author{ John F. Donoghue}
%\email[Email: ]{donoghue@physics.umass.edu}
\author{Basem Kamal El-Menoufi}
%\email[Email: ]{bmahmoud@physics.umass.edu}
\affiliation{Department of Physics,
University of Massachusetts\\
Amherst, MA  01003, USA}

\begin{abstract}
We discuss the derivation of the trace anomaly using a non-local effective action at one loop. This provides a simple and instructive form and emphasizes {\em infrared} physics. We then use this example to explore several of the properties of non-local actions, including displaying the action for the full non-local energy-momentum tensor. As an application, we show that the long-distance corrections at one loop lead to quantum violations of some classical consequences of the equivalence principle, for example producing a frequency dependence of the gravitational bending of light.
\end{abstract}
%\vspace{0.2 in}
%\end{titlepage}
%\setcounter{page}{0}
%\newpage
\maketitle
\section{Introduction}

We are used to dealing with local effective Lagrangians. However, one can also use non-local effective actions to summarize the one-loop predictions of a theory containing light or massless particles (see e.g. \cite{Gasser}). The non-locality occurs because light particles propagate a long distance within loop processes. In this paper, we explore some of the properties of such non-local effective actions in a simple context -  that of the energy momentum tensor in gauge theories with with massless particles.

One of the simplest and most instructive derivations of the QED trace anomaly is also one of the
least known. Let us present a quick treatment of this derivation, which we will then
explore in more detail in the body of this paper. In the massless limit, the classical electromagnetic action with charged matter is invariant under
the continuous rescaling
\begin{align}\label{scaletrans}
A_\mu(x) \to A^\prime_\mu(x^\prime) = \lambda^{-1} A_\mu(x),\quad  \psi(x)\to \psi^\prime(x^\prime) = \lambda^{-3/2}\psi(x),\quad  \phi(x)\to \phi^\prime(x^\prime) = \lambda^{-1} \phi(x) \  \ .
\end{align}
with $x'=\lambda x$. Associated with this symmetry is a scale or dilatation current\footnote{There are subtleties associated with the exact relation between the dilatation current and the energy-momentum tensor \cite{Polchinski} which we briefly discuss in Appendix A.}
\begin{equation}\label{Dcurrent}
J^\mu_{D} = x_\nu T^{\mu\nu}
\end{equation}
and the invariance of the action then leads to the tracelessness
of the energy momentum tensor
\begin{equation}\label{divD}
\partial_\mu J^\mu_{D} = T_\mu^{~\mu} = \frac{\partial { \cal \hat{L}_\lambda}}{\partial \lambda}\bigg|_{\lambda=1} = 0
\end{equation}
where ${\cal \hat{L}}_\lambda = \lambda^{4}{\cal L}(A^\prime,\psi^\prime, \phi^\prime)$ is independent of $\lambda$ when the action is scale invariant. With the symmetric energy momentum tensor for the photon,
\begin{align}\label{classicaleom}
T_{\mu\nu} = - F_{\mu\sigma} F_{\nu}^{\sigma} + \frac{1}{4} g_{\mu\nu} F_{\alpha\beta}F^{\alpha\beta} \ \
\end{align}
this property is readily apparent.

If we consider loops of the massless charged fields\footnote{All fields will be treated as massless in this paper. While there are no strictly massless charged particles, the results will apply at momentum transfer well above the particle mass. Moreover, these massless calculations are illustrative of other interesting situations, such as QCD or gravity, where strictly massless particles do appear.}, the vacuum polarization diagram
will contain a divergent piece which goes into
the renormalization of the electric charge. It also contains a $\ln q^2$ in momentum space, where $q_\mu$ refers to the
momentum of the photon. Rescaling the gauge field by the bare electric charge $A_\mu \to A_\mu/e_0$, we can write a one-loop
effective action describing both of these effects
\begin{equation}\label{quasilocal}
S = \int d^4x ~-\frac14 F_{\rho\sigma} \left[\frac{1}{e^2(\mu)} + b_i \ln \left({\Box}/{\mu^2}\right)\right]F^{\rho\sigma}
\end{equation}
where $b_i$ is the leading coefficient of the beta function, $b_s=1/(48\pi^2)$ for a charged scalar and $b_f=1/(12\pi^2)$ for a charged fermion, and $\Box = \partial^2$.

Under a scale transformation, we see that the
$\ln \Box$ term violates the scaling invariance since $\ln \Box \to  \ln \Box - \ln \lambda^2$.
From Eq. (\ref{divD}), we now infer that
\begin{equation}
\partial_\mu J_{D}^{\mu} = \frac{b_i}{2}F_{\rho\sigma}F^{\rho\sigma}   \ \ .
\end{equation}
After reverting to the usual definition of the field this
yields the usual form of the trace anomaly
\begin{equation}\label{anomop}
T_\mu^{~\mu}  = \frac{b_ie^2}{2}F_{\rho\sigma}F^{\rho\sigma}  \ \ .
\end{equation}

This derivation is instructive because it highlights the key physics - that the anomaly is related to the scale dependence of the running coupling,
which breaks the classical scale invariance. However, the procedure is also unusual in that the anomaly is associated
with an {\em infrared} effect, the $\ln q^2$ or $\ln \Box$ behavior. Most derivations and discussions of anomalies
emphasize the ultraviolet origin of the effect, either through regularization of the path integral or through the UV
properties of Feynman diagrams. Of course, the UV (the renormalization of the charge) and the IR (the $\ln q^2$) are
tied together when using dimensional regularization with massless fields, so there is not a contradiction. However, it
is satisfying to our effective field theory sensibilities to see a derivation that is insensitive to the UV regularization.
No matter how one regulates or modifies the high energy end of the theory (consistent with gauge invariance of course) the
infrared behavior and the trace anomaly will remain unaffected\footnote{There are also infrared derivations of the chiral anomaly \cite{Horejsi1}
and the trace anomaly \cite{Horejsi2, Giannotti} which make use of dispersion relations, with the integrand in the dispersive integral being dominated by
low energy contributions.}.

The Lagrangian of Eq. (\ref{quasilocal}) is written in {\em quasi-local} form, which we will explain in more detail below. The
$\ln \Box $ term is a shorthand for a non-local object
\begin{equation}
\langle x|\ln \left(\frac{\Box}{\mu^2}\right)|y\rangle \equiv L(x-y) = \int \frac{d^4q}{(2\pi)^4} e^{-iq\cdot (x-y)} \ln \left(\frac{-q^2}{\mu^2}\right)\ \ .
\label{logbox}
\end{equation}
However, under rescaling, this behaves in the same way as described above with a local term
\begin{equation}
L(x-y) \to \lambda^{-4}\left(L(x-y) - \ln \lambda^2 \delta^4(x-y)\right)
\end{equation}
yielding the same trace anomaly equation. It is well known that the anomaly does not follow from any local Lagrangian.
Here, we have seen that it does follow from the variation of a non-local Lagrangian.

As far as we know, this derivation was first sketched by Deser, Duff and Isham in a paper on gravitational conformal
anomalies \cite{DDI}. One can find echoes of it throughout the gravitational literature, for example in \cite{Barvinsky4, Duff1, Schwimmer, Armillis1, Deser2, Duff2, Wald}, which is surely an
incomplete list. The local anomaly itself has been thoroughly discussed in the literature and we have little new to add. However, our objective in this paper is two-fold. The first concerns the connection of anomalies to non-local effective actions which is not regularly discussed in the gauge
theory literature. Our purpose here will be to give a thorough discussion of this non-local effect for QED and to use this simple
example to make a concrete exploration of non-local effective actions. A second goal is to discuss the extra novel features when we include the gravitational
coupling in the non-local actions. This provides a simple example of non-local gravitational actions, which is an interesting but more complicated subject.

After finding a local trace anomaly from a non-local action, it is natural to consider the full energy-momentum tensor which yields the
appropriate trace. Due to the propagation of massless particles in the loop, it will also be a non-local object. To our knowledge, this object has not been constructed before in the literature. This step is indeed important if one wants to fully understand the phenomenology of the trace anomaly. We will construct this object for a charged scalar field in the loop and later display the result for fermions by consulting the matrix element calculation of \cite{Berends, Milton}. An extra motivation for using a charged scalar is that, unlike fermions, the scalar's minimally coupled action is {\em not} conformally invariant. This provides an interesting insight into the connection between conformal/scale invariance and the anomaly.
Our non-local form also has several interesting properties, which we discuss.

In regard to gravity, we also provide a {\em partial} non-linear completion of the perturbative result using the gravitational curvatures, although we
reserve a detailed discussion of this aspect to a companion publication \cite{Donoghue}. Our result for the traceful part of the energy-momentum tensor can be
obtained by varying a covariant action
\begin{align}
T^{anom.}_{\mu\nu} = \left(\frac{2}{\sqrt{g}} \frac{\delta \Gamma[g,A]}{\delta g^{\mu\nu}}\right)_{g=\eta}
\end{align}
where
\begin{align}
\Gamma[g,A] = \int d^4x \, \sqrt{g} \left(n_R F_{\rho\sigma}F^{\rho\sigma}\frac{1}{\Box} R + n_C F^{\rho\sigma} F^{\gamma}_{\, \, \lambda} \frac{1}{\Box}C_{\rho\sigma\gamma}^{\quad \lambda} \right)  \ \ .
\label{nonlinearcompletionintro}
\end{align}
Here, $C_{\rho\sigma\gamma}^{\quad \lambda}$ is the Weyl tensor and $\Box$ is the covariant d' Alembertian. We will find that the first coefficient is determined
by the beta functions of fermions or bosons
\begin{align}
n^{(s,f)}_R = - \frac{\beta^{(s,f)}}{12 e}
\end{align}
while the last coefficient is not related to the beta functions and does not contribute to the trace. Note the $1/\Box$ pole which appears in the action which is required by direct calculation of the effective action.

Since the energy momentum-tensor describes the coupling of photons to gravity, we also look at the scattering of a photon by the gravitational field of a massive object. The quantum corrections carry an extra energy dependence that leads to violations of some of the predictions of classical general relativity. For example, the equivalence principle requires that the bending of light is the same for photons of all energies. We show that this is no longer the case when non-local loop effects
are present. We should expect that this quantum violation of the equivalence principle
should be a general phenomenon, as noted in \cite{Bohr}. Within our calculation it could be described as a ``tidal'' effect since the photon's coupling is no longer a local object but
samples the gravitational field over a long distance through quantum loops of massless particles. Quantum mechanics does this in general by producing spatial
non-localization and our example provides a non-trivial demonstration of this property\footnote{Of course, since all charged particles in Nature have mass, the results will only be applicable in the real world for photons with energies well above the electron mass.}.

\section{The background field method and the non-local effective action}

Here we give a brief derivation of the non-local effective action using the background field method. The classical action for QED coupled to a charged field reads
\begin{align}
S = S_{EM} + \int d^4x \, (D_{\mu}\phi)^{\star} D^{\mu}\phi
\end{align}
where
\begin{align}
D_{\mu} \phi = (\partial_{\mu} + i e_0 A_{\mu} )\phi, \quad S_{EM} =\int d^4x \, - \frac{1}{4} F_{\mu\nu}F^{\mu\nu}
\end{align}
and $e_0$ is the bare electric charge.

The one loop effective action is obtained by integrating out the charged scalar field
\begin{align}\label{PI}
\nonumber
\Gamma [A] &= \frac{1}{e_0^2} S_{EM} - i \ln \left(\int \mathcal{D} \phi^{\star}\, \mathcal{D} \phi \, e^{iS} \right)\\
&=\frac{1}{e_0^2} S_{EM} + i \ln \left(\text{Det}\, D^2 \right)
\end{align}
where we rescaled the gauge field. The operator reads
\begin{align}
D^2 = \Box + i(\partial \cdot A) + 2i A^\mu \partial_\mu -  A^2 \ \ .
\end{align}
In perturbation theory we can expand the logarithm in powers of the interaction
\begin{equation}
\text{ln} \left(\text{Det}\, D^2 \right) = \text{Tr} \left(\frac{1}{\Box}\, v - \frac12 \, \frac{1}{\Box} \, v \frac{1}{\Box} \, v  +....\right) + \text{const.}
\end{equation}
where
\begin{align}
v = i (\partial \cdot A) + 2i A^\mu \partial_\mu -  A^2 \ \ .
\end{align}
Introducing position-space eigenstates such that
\begin{align}\label{Fprop}
\langle x| \frac{1}{\Box}|y\rangle =i \Delta_F(x-y)
\end{align}
and using dimensional regularization, we have that $\Delta_F(0) = 0$, and hence the first term in the expansion vanishes. Integrating by parts to place the derivatives on the propagators and noting that the latter is a function of the geodesic distance $|x-y|$, we find the order-$A^2$ contribution
\begin{align}
\Gamma[A] = \frac{1}{e_0^2} S_{EM} + i \int d^Dx \, d^Dy  \,  A^\mu(x) M_{\mu\nu}(x-y) A^\nu(y)
\end{align}
and
\begin{align}
M_{\mu\nu}(x-y) = \partial_\mu \, \Delta_F(x-y) \partial_\nu \, \Delta_F(x-y) - \Delta_F(x-y) \partial_\nu \partial_\mu \, \Delta_F(x-y)
\end{align}
and all derivatives act on $x$. By Fourier transforming and using standard manipulations in momentum space, one obtains the following
relations
\begin{eqnarray}
\nonumber
\Delta_F(x)\partial_\mu \Delta_F(x) &=& \frac12 \partial_\mu \Delta^2_F(x) \nonumber \\
\Delta_F(x)\partial_\mu \partial_\nu \Delta_F(x) &=& \left[d\partial_\mu \partial_\nu - g_{\mu\nu} \Box\right] \frac{\Delta^2_F(x)}{4(d-1)} \nonumber \\
\partial_\mu \Delta_F(x)\partial_\nu \Delta_F(x)&=& \left[(d-2)\partial_\mu \partial_\nu + g_{\mu\nu} \Box\right] \frac{\Delta^2_F(x)}{4(d-1)} \ \ .
\end{eqnarray}
These combine to produce a tensor
\begin{equation}
M_{\mu\nu}(x-y) = \left[ g_{\mu\nu} \Box - \partial_\mu \partial_\nu \right]  \frac{\Delta^2_F(x-y)}{2(d-1)}
\end{equation}
which is conserved in any dimension. Converting one $x$-derivative back to one with respect to $y$ and integrating by parts
we convert the result to a manifestly gauge invariant form
\begin{equation}
\Gamma[A] = \frac{1}{e_0^2} S_{EM} - i \int d^Dx d^Dy \,  F_{\mu\nu}(x) \left[ \frac{\Delta^2_F(x-y)}{4(d-1)}   \right] F^{\mu\nu}(y) \ \ .
\end{equation}
We can represent the squared propagator by a Fourier transformation
\begin{align}
\Delta^2_F(x-y) = -\int \frac{d^Dq}{(2\pi)^D} e^{-iq(x-y)} I_2(q)
\end{align}
where $I_2(q)$ is the scalar bubble function which reads
\begin{align}
I_2(q) = \frac{i}{16\pi^2} \left[\frac{1}{\bar{\epsilon}} - \ln\left(\frac{-q^2}{\mu^2}\right)\right], \quad \frac{1}{\bar{\epsilon}} = \frac{1}{\epsilon} - \gamma + \ln 4\pi \ \ .
\end{align}
with $\epsilon = (4-D)/2$. Now it is easy to renormalize the electric charge\footnote{Note that since $[1/(d-1)]1/\epsilon=1/(3\epsilon) +2/3 $, there is an extra constant factor of 2/3 when using modified Minimal Subtraction renormalization. This constant is irrelevant for our purposes and we do not display it.} and hence express the 4D effective action in a quasi-local form
\begin{align}
\Gamma[A] = \int d^4x ~-\frac{1}{4} F_{\mu\nu} \left[\frac{1}{e^2(\mu)} + b_i \ln \left(\frac{\Box}{\mu^2}\right)\right]F^{\mu\nu}
\end{align}
where we find for the scalar loop (and by analogy for the fermion loop)
\begin{align}
b_s = \frac{1}{48\pi^2}, \quad b_f = \frac{1}{12\pi^2} \ \ .
\end{align}

\section{Including the energy momentum tensor in the effective action}

The trace of the energy momentum tensor is a local object. What about the full energy-momentum tensor $T_{\mu\nu}$ itself? One might try following the conventional procedure by employing the translation invariance of the quasi-local action in Eq. (\ref{quasilocal}) to find $T_{\mu\nu}$, but the non-local term renders this task impossible. One elegant pathway is to compute the effective action in curved space from which we can identify the energy momentum tensor through the relation
\begin{align}\label{emt}
\delta \Gamma[g,A] = \frac{1}{2} \int d^4x \,\sqrt{g} \, \delta g^{\mu\nu} \, T_{\mu\nu} \ \ .
\end{align}
Hence we are interested in the non-local effective action including gravity. Of course we cannot complete this program for an arbitrary gravitational field. However it is sufficient to use perturbation theory if our aim is just the flat space result. Moreover, as we show in Sect. (6), perturbation theory can be used to propose a non-linear completion of the effective action apart from subtleties that we address in \cite{Donoghue}. We perform the computation for bosons and consult \cite{Berends, Milton} to read off the result for fermions. The starting point is the action
\begin{align}\label{scalaraction}
S = S_{EM} + \int d^Dx \sqrt{g} \big[g^{\mu\nu}(D_{\mu}\phi)^{\star}(D_{\nu}\phi) - \xi \phi^{\star}\phi R \big]
\end{align}
where all derivative operators are covariant.

We have included the $\xi\phi^{\star}\phi R $ coupling, with $\xi=0$ being minimally coupled and $\xi =1/6$ being conformally coupled, in order to separately
follow scale and conformal symmetry. For $\xi=1/6$ the above action is invariant under {\em local} Weyl transformations, i.e. conformal transformations. Namely,
\begin{align}\label{weyltrans}
g_{\mu\nu} \rightarrow e^{2\sigma(x)} g_{\mu\nu}, \quad \phi \rightarrow e^{-\sigma(x)} \phi, \quad A_{\mu} \rightarrow A_{\mu} \ \ .
\end{align}

On the other hand, the minimally coupled action is invariant only under scale transformations. The scalar field energy-momentum tensor
\begin{align}
T_{\mu\nu} = (\partial_{\mu}\phi)^{\star}(\partial_{\nu}\phi)+ (\partial_{\nu}\phi)^{\star}(\partial_{\mu}\phi) - g_{\mu\nu}(\partial_{\lambda}\phi)^{\star}(\partial^\lambda \phi) + 2\xi (g_{\mu\nu}\Box -\partial_\mu\partial_\nu)\phi^\star\phi - 2\xi(R_{\mu\nu}-\frac12 g_{\mu\nu}R)\phi^\star\phi
\end{align}
is traceless only for $ \xi =1/6$.
For future reference, we point out that the trace of the energy-momentum tensor could be directly determined by performing a conformal transformation and then varying the action with respect to $\sigma$, namely
\begin{align}\label{sigmavar}
\delta_{\sigma} S = - \int d^4x\, \sigma \, T_{\mu}^{\, \mu} \ \ .
\end{align}

Turning to our calculation, we start by performing the path-integral which yields Eq. (\ref{PI}) but with the curved space operator
\begin{align}
D^2 = \sqrt{g} \left(\nabla^{\mu}\nabla_{\mu} + 2i A^{\mu} \partial_{\mu} + i \nabla_{\mu} A^{\mu} - A_{\mu}A^{\mu} + \xi R\right) \ \ .
\end{align}
The perturbative calculation is set up by expanding the metric around flat space
\begin{align}
g_{\mu\nu} = \eta_{\mu\nu} + h_{\mu\nu}
\end{align}
and all other geometric quantities accordingly. From Eqn. (\ref{emt}), it suffices to compute the effective action linear in the perturbation $h_{\mu\nu}$ up to terms quadratic in the gauge field. There exist three diagrams which contribute at this order, a triangle Fig. [\ref{tr}] and two bubble-like diagrams Fig. [\ref{bb}]. We evaluate the effective action {\em on-shell}, and thus impose both on-shellness of external photons $p^2 = p^{\prime 2} =0$ and transversality $p\cdot A(p)=p^{\prime}\cdot A(p^{\prime})=0$.

\begin{figure*}
\centering
\epsfig {file=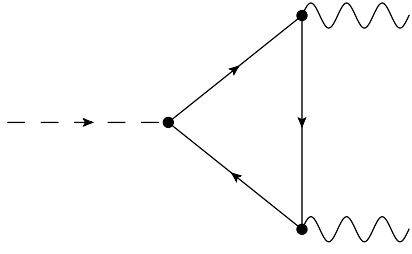,width=0.3\linewidth,clip=}
\caption{Triangle diagram.}
\label{tr}
\end{figure*}

\begin{figure*}
\centering
\begin{tabular}{cc}
\epsfig {file=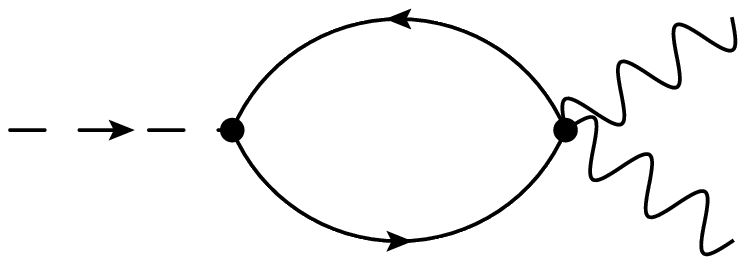,width=0.3\linewidth,clip=}& \quad\quad\quad
\epsfig {file=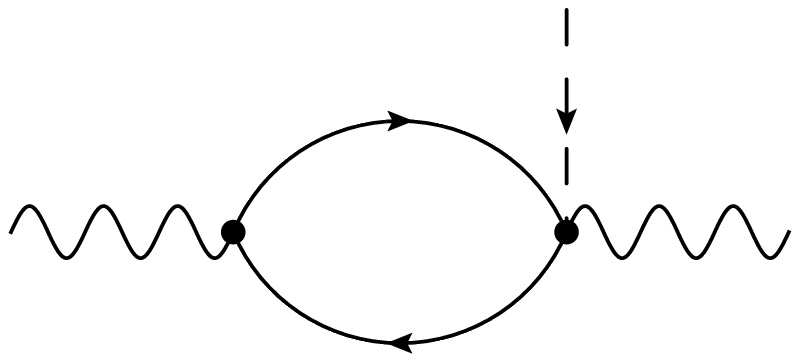,width=0.3\linewidth,clip=}
\end{tabular}
\caption{Bubble diagrams.}
\label{bb}
\end{figure*}
The calculation is performed using the Passarino-Veltman (P-V) reduction technique \cite{Passarino}, the details of which are included in an appendix.  The result of the triangle diagram is
\begin{align}
\mathcal{T} = \int_{p} \int_{p^{\prime}} \tilde{h}^{\mu\nu}(-q)\,  \tilde{A}^{\alpha}(p) \, \tilde{A}^{\beta}(-p^{\prime})\, \mathcal{P}^T_{\mu\nu, \alpha\beta}
\end{align}
where
\begin{align}
\nonumber
\mathcal{P}^T_{\mu\nu, \alpha\beta} &= [4 H + B q^2]\eta_{\mu\nu}\eta_{\alpha\beta} + 4 H \left(\eta_{\mu\alpha}\eta_{\nu\beta} + \eta_{\mu\beta}\eta_{\nu\alpha}\right) + [4 I - 4 J + C q^2 - D q^2] \eta_{\mu\nu} p_{\alpha}^{\prime} p_{\beta} + [4 I + 4 E + B] Q_{\mu}Q_{\nu} \eta_{\alpha\beta} \\\nonumber
&+ [4 J - B] q_{\mu}q_{\nu} \eta_{\alpha\beta} + [4 K + 4 F + C -4 M - 4 G - D] Q_{\mu}Q_{\nu}p_{\alpha}^{\prime} p_{\beta} + [4 M - C - 4 L + D] q_{\mu}q_{\nu}p_{\alpha}^{\prime} p_{\beta} \\
&+ [4 I + 2 E - 4J] (p_{\alpha}^{\prime} p_{\mu} \eta_{\nu\beta} + p_{\mu}^{\prime} p_{\beta} \eta_{\nu\alpha} + p_{\alpha}^{\prime} p_{\nu} \eta_{\mu\beta} + p_{\nu}^{\prime} p_{\beta} \eta_{\mu\alpha}) - 4 \xi (q_{\mu} q_{\nu} - q^2 \eta_{\mu\nu}) (B\eta_{\alpha\beta} + (C-D)p^{\prime}_\alpha p_{\beta}) \ \ .
\end{align}
Here the various coefficients are the result of performing the momentum integration - these are given in the appendix. The first of the bubble diagrams reads
\begin{align}
\mathcal{B}_1=\int_{p} \int_{p^{\prime}} \tilde{h}^{\mu\nu}(-q)\,  \tilde{A}^{\alpha}(p)  \tilde{A}^{\beta}(-p^{\prime})\, \mathcal{P}^{B_1}_{\mu\nu,\alpha\beta}
\end{align}
where
\begin{align}
\mathcal{P}^{B_1}_{\mu\nu,\alpha\beta} = \left[\frac{D-2}{4(D-1)} - \xi \right] (q^2 \eta_{\mu\nu} - q_{\mu} q_{\nu})\eta_{\alpha\beta} I_2 (q) \ \ .
\end{align}
The last diagram reads
\begin{align}
\mathcal{B}_2 = 2 \int_{p} \int_{p^{\prime}} \tilde{h}^{\mu\nu}(-q)\,  \tilde{A}^{\alpha}(p) \, \tilde{A}^{\beta}(-p^{\prime})\, \mathcal{P}^{B_2}_{\mu\nu, \alpha\beta}
\end{align}
where
\begin{align}
\mathcal{P}^{B_2}_{\mu\nu, \alpha\beta} = \frac{1}{2} \left(\eta_{\beta\mu}p_{\nu}p_\alpha + \eta_{\beta\nu}p_{\mu}p_\alpha -\frac{1}{2}\eta_{\mu\nu}p_{\beta}p_\alpha\right)I_2(p) -\frac{D}{4(D-1)}\left(\eta_{\beta\mu}p_\nu p_\alpha-\eta_{\beta\nu}p_\mu p_\alpha +\frac{1}{2}\eta_{\mu\nu}p_\alpha p_\beta\right)I_2(q) \ \ .
\end{align}
This last diagram vanishes simply due to the condition $p\cdot\tilde{A}(p)=0$.

Combining the three diagrams we find that to this order in perturbation theory the effective action reads
\begin{align}
\Gamma[g,A] = S_{EM} - i \int_{p} \int_{p^{\prime}} \tilde{h}^{\mu\nu}(-q)\,  \tilde{A}^{\alpha}(p) \, \tilde{A}^{\beta}(-p^{\prime})\, \mathcal{M}_{\mu\nu,\alpha\beta}
\end{align}
where
\begin{align}
\nonumber
\mathcal{M}_{\mu\nu,\alpha\beta}&=\mathcal{P}^T_{\mu\nu, \alpha\beta}-\mathcal{P}^{B1}_{\mu\nu, \alpha\beta}\\
&=\left(\frac{1}{12}\mathcal{M}^0_{\mu\nu,\alpha\beta} + \frac{1}{q^2} \left[a Q_{\mu}Q_{\nu} \left(p^{\prime}_{\alpha}p_{\beta}-p\cdot p^{\prime}\eta_{\alpha\beta}\right)+ b \left(q_{\mu}q_{\nu} - q^2 \eta_{\mu\nu}\right)\left(p^{\prime}_{\alpha}p_{\beta}-p\cdot p^{\prime}\eta_{\alpha\beta}\right)\right]\right)I_2(q)
\end{align}
and
\begin{align}
a=-\frac{1}{24}(D-4) , \quad b=\left[\frac{5}{24}-\xi\right](D-4)
\label{coefficients}
\end{align}
and $\mathcal{M}^0_{\mu\nu,\alpha\beta}$ is the lowest order photon energy momentum matrix element
\begin{align}
\nonumber
\mathcal{M}^0_{\mu\nu,\alpha\beta}&=p_{\mu}^{\prime} p_{\nu} \eta_{\alpha\beta} + p_{\mu} p^{\prime}_{\nu} \eta_{\alpha\beta} + \eta_{\mu\nu} p^{\prime}_{\alpha} p_{\beta} - p_{\mu} p^{\prime}_{\alpha} \eta_{\nu\beta} - p_{\mu}^{\prime} p_{\beta} \eta_{\alpha\nu} - p_{\nu} p^{\prime}_{\alpha} \eta_{\mu\beta} \\
&- p_{\nu}^{\prime} p_{\beta} \eta_{\alpha\mu} + p \cdot p^{\prime} (\eta_{\mu\alpha} \eta_{\beta\nu} + \eta_{\mu\beta} \eta_{\nu\alpha} - \eta_{\mu\nu} \eta_{\alpha\beta}) \ \ .
\end{align}
We have the limit $D=4$ in all terms except for those of Eq. \ref{coefficients}.

There are a couple of interesting calculational features in this computation. One is that although we are calculating a triangle diagram, the scalar triangle integral
\begin{align}
I_3(p,p') = \int \frac{d^Dk}{(2\pi)^D}   \frac{1}{(l^2 + i0)((k+p^2)+i0) ((k + p^{\prime})^2 + i0)}
\end{align}
does not appear in the result. The above integral is infrared divergent, and thus despite the massless loops the on-shell conditions yielded an infrared finite effective action up to this order in perturbation theory. The P-V reduction has expressed all of the integrals in terms of the bubble integral and the answer only
contains
\begin{align}\label{bubble}
I_2(q) = \int \frac{d^Dk}{(2\pi)^D}   \frac{1}{(k^2 + i0)((k+q)^2 + i0)}=\frac{i}{16\pi^2} \left[\frac{1}{\bar{\epsilon}} - \ln\left(\frac{-q^2}{\mu^2}\right)\right]
\end{align}
with $\frac{1}{\bar{\epsilon}} = \frac{1}{\epsilon} - \gamma + \ln 4\pi$. Also interesting is that the bubble integral as a function of an external momenta
\begin{align}\label{infraredbubble}
I_2(p^2=\lambda^2)&= \int \frac{d^Dk}{(2\pi)^D}   \frac{1}{(k^2+i0)((k+p)^2 + i0)} = \frac{i}{16\pi^2} \left[\frac{1}{\bar{\epsilon}} - \ln\left(\frac{-\lambda^2}{\mu^2}\right)\right]
\end{align}
does not appear in the answer. In doing the P-V reduction shown in the appendix, we kept the off-shell condition $p^2=p'^2= \lambda^2$ in potentially divergent contributions in order to regulate the infrared aspects of the integrals, and inspection of these integrals shows $I_2(\lambda^2)$ occurring frequently. However, all such terms drop out of the final result.

\subsection{Renormalization}
It is expected that the divergent part of the effective action is proportional to $S_{EM}$ which reads in momentum space
\begin{align}
S_{EM} = \frac{1}{4e_0^2} \int_{p} \int_{p^{\prime}} \, \tilde{h}^{\mu\nu}(-q)\,  \tilde{A}^{\alpha}(p) \, \tilde{A}^{\beta}(-p^{\prime}) \mathcal{M}^0_{\mu\nu,\alpha\beta}
\end{align}
and $e_0$ is the bare electric charge.
As usual, the bare electric charge is replaced by its renormalized counterpart via
\begin{align}
e_0 = \mu^{\epsilon} \, Z^{-1/2}_3 \, e \ \ .
\end{align}
Working in the modified $MS$-scheme the renormalization constant is easily determined to be
\begin{align}
Z_3 = 1 - \frac{e^2}{48 \pi^2 \, \bar{\epsilon}} \ \ .
\end{align}
It is now easy to determine the beta function from the RGE
\begin{align}
\beta^s(e) = \frac{e^3}{48 \pi^2} \ \ .
\end{align}
After renormalization, we pass to the limit $D=4$ and write down the renormalized effective action
\begin{align}\label{renea}
\Gamma^{\text{ren}}[g,A] = \frac{1}{4} \int_{p} \int_{p^{\prime}}\, \tilde{h}^{\mu\nu}(-q)\,  \tilde{A}^{\alpha}(p) \, \tilde{A}^{\beta}(-p^{\prime}) \left[\left(\frac{1}{e^2(\mu)} - \frac{1}{48 \pi^2}\ln\left(\frac{-q^2}{\mu^2}\right)\right) \mathcal{M}^0_{\mu\nu,\alpha\beta} + \mathcal{M}^s_{\mu\nu,\alpha\beta}\right]
\end{align}
where we identified the finite tensor for the charged scalar leaving the value of the conformal coupling arbitrary
\begin{align}
\mathcal{M}^s_{\mu\nu,\alpha\beta}(\xi) = \frac{1}{48 \pi^2 q^2}\left(Q_{\mu}Q_{\nu} - (5-24\xi)( q_{\mu} q_{\nu} - q^2 \eta_{\mu\nu}) \right) \left(p_\alpha^{\prime} p_{\beta} - p \cdot p^{\prime} \eta_{\alpha\beta} \right) \ \ .
\end{align}

We see that only for $\xi=1/6$ does the photon's energy momentum tensor have the expected trace relation. The lack of Weyl invariance in the scalar sector when
$\xi\ne 1/6$ carries over to the photon interaction and modifies the trace.  As we show below, this feature is not present for fermions since the classical theory is Weyl invariant. On the other hand, it is satisfying to observe that, using the beta function, the renormalized effective action is indeed scale-independent.

\subsection{Fermions and non-universality}

At this stage, it is quite straightforward to read off the result for fermions from the matrix-element computation of \cite{Berends}
\begin{align}
\Gamma^{\text{ren}}[g,A] = \frac{1}{4} \int_{p} \int_{p^{\prime}}\, \tilde{h}^{\mu\nu}(-q)\,  \tilde{A}^{\alpha}(p) \, \tilde{A}^{\beta}(-p^{\prime}) \left[\left(\frac{1}{e^2} - \frac{1}{12 \pi^2}\ln\left(\frac{-q^2}{\mu^2}\right)\right) \mathcal{M}^0_{\mu\nu,\alpha\beta} + \mathcal{M}^f_{\mu\nu,\alpha\beta}\right]
\end{align}
where the finite tensor now becomes
\begin{align}
\mathcal{M}^f_{\mu\nu,\alpha\beta} = \frac{1}{24 \pi^2 q^2}\left(-Q_{\mu}Q_{\nu} - q_{\mu} q_{\nu} + q^2 \eta_{\mu\nu} \right) \left(p_\alpha^{\prime} p_{\beta} - p \cdot p^{\prime} \eta_{\alpha\beta} \right) \ \ .
\end{align}
We also find the fermionic beta function
\begin{align}
\beta^f(e)=\frac{e^3}{12\pi^2} \ \ .
\end{align}
An interesting aspect of this result is the {\em non-universality} of the structure of the finite tensor which is responsible for the anomalous trace. However, we will show below that the trace of this tensor reproduces the correct anomaly for both bosons and fermions.

\subsection{Position space effective action}

Let us collect these calculations into a position space effective action. After integrating out the massless charged particle, it has the general structure
\begin{align}
\Gamma[A,h] = \frac{1}{e^2(\mu)} S_{EM}[A,h] + \Gamma^{(0)}[A] + \Gamma^{(1)}[A,h]
\end{align}
where
\begin{align}
S_{EM}[A,h] = -\frac14 \int d^4x \left( F_{\mu\nu} F^{\mu\nu} + 2  \, h^{\mu\nu}\, T_{\mu\nu}^{cl} \right)
\end{align}
with $T_{\mu\nu}^{cl}(x)$ given by Eq. (\ref{classicaleom}) and $\Gamma^{(0)}[A]$ being the non-local piece in Eq. (\ref{quasilocal}). The loop corrections linear in $h_{\mu\nu}$ are contained in $\Gamma^{(1)}[A,h]$. Written in quasi-local form, it has the structure \footnote{From now onwards, we use $\xi=1/6$.}
\begin{align}\label{NLactions}
\Gamma^{(1)}[A,h]=-\frac12 \int d^4x \, h^{\mu\nu}\left[ b_s \log \left(\frac{\Box}{\mu^2}\right)T_{\mu\nu}^{cl} - \frac{b_s}{2}\frac{1}{\Box} \tilde{T}^{s}_{\mu\nu} \right]
\end{align}
for conformally coupled scalars, where $b_s$ is the beta function coefficient and $\tilde{T}^{s}_{\mu\nu} $ is the operator
\begin{align}
\tilde{T}^s_{\mu\nu} = 2 \partial_{\mu}F_{\alpha\beta}\partial_{\nu}F^{\alpha\beta} - \eta_{\mu\nu}\partial_{\lambda}F_{\alpha\beta}\partial^{\lambda}F^{\alpha\beta} \ \ .
\end{align}
For fermions, the structure is similar
\begin{align}\label{NLactionf}
\Gamma^{(1)}[A,h]= -\frac12 \int d^4x \, h^{\mu\nu}\left[ b_f \log \left(\frac{\Box}{\mu^2}\right)T_{\mu\nu}^{cl} - \frac{b_f}{2}\frac{1}{\Box} \tilde{T}^{f}_{\mu\nu} \right]
\end{align}
except now $\tilde{T}^{f}_{\mu\nu}$ is a different operator
\begin{align}
\tilde{T}^f_{\mu\nu} &= - F_{\alpha\beta} \partial_{\mu} \partial_{\nu}F^{\alpha\beta} - \frac12 \eta_{\mu\nu}\partial_{\lambda}F_{\alpha\beta}\partial^{\lambda}F^{\alpha\beta} \ \ .
\end{align}

Both of these are genuine non-local actions. To display the non-locality we recall that the $\log \Box$ factor is to be interpreted as in Eq. (\ref{logbox}), and equivalently the $1/\Box$ term is the representation of the Feynman propagator as in Eq. (\ref{Fprop}). \footnote{When using the in-in formalism, the causal prescription for the $\ln \Box$ piece was computed in \cite{Basem} and evidently the $1/\Box$ would be the retarded propagator.}
Then the explicitly non-local form reads
\begin{align}
\Gamma^{(1)}[A,h]=-\frac12 \int d^4x \, h^{\mu\nu}(x) \int d^4 y \,\left[ b_i \, L(x-y)T_{\mu\nu}^{cl}(y) -i \frac{b_i}{2} \, \Delta_F(x-y) \tilde{T}^{i}_{\mu\nu} (y) \right], \quad i=s,f \ \ .
\end{align}
We see both a logarithmic non-locality and a mass-less pole non-locality.

From Eq. (\ref{emt}), one can readily obtain the energy momentum tensor itself from these expressions. In doing so, we rescale the photon field by a factor of $e(\mu)$ in order to obtain the conventional normalization. The result is given by the
non-local object
\begin{align}\label{nlems}
T^i_{\mu\nu}(x) =T_{\mu\nu}^{cl}(x) - e^2 b_i\int d^4y \, \left [{L}(x-y) T_{\mu\nu}^{cl}(y) + \frac{i}{2}\, \Delta_{F}(x-y) \tilde{T}^i_{\mu\nu}(y) \right], \quad i=s,f \ \ .
\end{align}
Note that this form does contain a dependence on the scale $\mu$ within the logarithm.
Using the on-shell condition $\Box A_{\mu} = 0$ we have that
\begin{align}\label{trick}
\partial_{\lambda}F_{\alpha\beta}\partial^{\lambda}F^{\alpha\beta} = \frac{1}{2} \Box \left(F_{\mu\nu}F^{\mu\nu}\right)
\end{align}
and thus one can easily verify that the above tensor reproduces the correct trace anomaly. Moreover, to show that it is conserved one merely notices that both non-local functions are functions of the geodesic distance and hence convert derivatives to be with respect to the $y$ variable and then uses integration by parts. Eq. (\ref{nlems}) is one of the main results of this paper.

One can gain some insight into this structure if one decomposes the boson and fermion tensors into a universal term which yields the proper trace and a non-universal term that is traceless. Here we find
\begin{align}
\tilde{T}^i_{\mu\nu} = a_1^i A_{\mu\nu} + a_2^i S_{\mu\nu}, \quad i=s,f
\end{align}
where
\begin{align}
A_{\mu\nu} &= \partial_\mu F_{\alpha\beta} \partial_\nu F^{\alpha\beta} + F_{\alpha\beta}\partial_\mu \partial_\nu F^{\alpha\beta} - \eta_{\mu\nu} \partial_\lambda F_{\alpha\beta} \partial^\lambda F^{\alpha\beta}\\
S_{\mu\nu} &= 4 \partial_\mu F_{\alpha\beta} \partial_\nu F^{\alpha\beta} - 2 F_{\alpha\beta}\partial_\mu \partial_\nu F^{\alpha\beta} - \eta_{\mu\nu} \partial_\lambda F_{\alpha\beta} \partial^\lambda F^{\alpha\beta}
\end{align}
and
\begin{align}
a_1^s = a_1^f = \frac23 , \quad a_2^s = \frac13 , \quad a_2^f =- \frac16 \ \ .
\end{align}
The trace of $A_{\mu\nu}$ gives the anomaly, while $S_{\mu\nu}$ is traceless. There is of course an ambiguity in any such decomposition - one can add any traceless tensor to $A_{\mu\nu}$ while subtracting it from $S_{\mu\nu}$. We have chosen the linear combinations to match the nonlinear completion that we will display in Sect. (6), such that $A_{\mu\nu}$ corresponds to the $F^2(1/\Box) R$ term and $S_{\mu\nu}$ to the $F^2(1/\Box)C$ term.

\section{Conformal and scaling properties of the effective action}

In the one loop effective action, we have found two terms that are proportional to the beta function coefficient, $b_i$. These can be referred to as the $\ln \Box$ term and the $1/\Box$ term. We will see that both of them are required, but by somewhat different scale symmetry transformations. As we will describe below, the $\ln \Box$ responds directly to dilations while the $1/ \Box$ responds to conformal transformations. The existence of both allows us to relate the two symmetries in this context. The $\ln \Box$ behavior and the $1/ \Box$ behavior are much discussed in the literature. For example, Deser and Schwimmer \cite{Schwimmer} refer to the $\ln \Box$ terms as Type B anomalies and $1/ \Box$ as Type A. It is interesting that both types emerge in this calculation. The $1/ \Box$ terms are also associated with the Riegert anomaly action \cite{Riegert}, which will be commented on in Sect. (6).

Let us now discuss the dichotomy between scaling and conformal symmetry breaking in the effective action constructed in the previous section. The scaling behavior of $\Gamma^{(0)}[A]$ was discussed in the introduction. Before we repeat the same exercise for $\Gamma^{(1)}[A,h]$, we note that since $h_{\mu\nu}$ has a mass dimension zero, it has a vanishing scaling dimension. Accordingly, under a scale transformation the 1-loop EA transforms as follows
\begin{align}
\Gamma^{(1)}[A,h] \rightarrow \Gamma^{(1)}[A,h] + \frac{b_i}{2} \int d^4x \, h^{\mu\nu}  \left[ \log \lambda^2 T^{cl}_{\mu\nu} \right] \ \ .
\end{align}
Using Eq. (\ref{divD}) and taking $\Gamma^{(0)}[A]$ into account as well, we find
\begin{align}
T_{\mu}^{\, \mu} = \frac{b_i}{2} \left(\eta^{\mu\alpha} \eta^{\nu\beta} F_{\mu\nu} F_{\alpha\beta} + 2 h^{\mu\nu} T^{cl}_{\mu\nu} \right)
\end{align}
which is indeed the desired anomalous operator expanded around flat space.

Notice in particular the feature that when performing this rescaling, the $1/\Box$ portion of the answer is scale invariant. However, when forming the energy momentum tensor, it is precisely the $1/\Box$ part that yields the {\em traceful} contribution to the energy-momentum tensor. To explain this, we need to understand the violation of conformal symmetry present in the effective action. Once again, we need to determine the transformation properties of the metric perturbation $h_{\mu\nu}$. This is best achieved by linearizing the classical action and performing an infinitismal conformal transformation, namely
\begin{align}
g_{\mu\nu} \rightarrow (1 + 2 \sigma) g_{\mu\nu} \ \ .
\end{align}
This allows to read off the transformation of the metric perturbation
\begin{align}
h_{\mu\nu} \rightarrow h_{\mu\nu} + 2 \sigma \eta_{\mu\nu}
\end{align}
One can readily check that the linearized action of Eq. (\ref{scalaraction}) is indeed invariant under the above transformation provided $\phi \rightarrow (1-\sigma)\phi$. Both $S_{EM}[A,h]$ and $\Gamma^{(0)}[A]$ are invariant. Moreover,
\begin{align}
\Gamma^{(1)}[A,h] \rightarrow \Gamma^{(1)}[A,h] - b_i \int d^4x \, \sigma \frac{1}{\Box} \left(\partial_{\lambda}F_{\mu\nu} \partial^{\lambda}F^{\mu\nu} \right) \ \ .
\end{align}
By using Eqs. (\ref{trick}) and (\ref{sigmavar}), one reproduces the flat space limit of the anomalous operator
\begin{align}
T_{\mu}^{\, \mu} = \frac{b_i}{2} \eta^{\mu\alpha} \eta^{\nu\beta} F_{\mu\nu} F_{\alpha\beta}
\end{align}

We have seen that when expanding to first order around flat space, two terms arise which are both related to the anomaly. When forming the energy momentum tensor, the log term multiplies the classical energy momentum tensor and hence is itself traceless. However under scale transformations the log produces an anomaly which combines with the lowest order piece in the proper way. On the other hand, conformal transformations directly produce the trace of the energy-momentum tensor, and this is manifest in the $1/\Box$ term of the one-loop result.

\section{The on-shell energy-momentum matrix element at one loop}

For completeness, let us display the matrix element of the energy momentum tensor found in the previous section. The energy momentum tensor for on-shell photons has the general form
\begin{align}
\langle \gamma(p')|T_{\mu\nu}|\gamma(p)\rangle &= \epsilon_\beta^*(p')\epsilon_\alpha(p)\left[  \mathcal{M}^0_{\mu\nu,\alpha\beta}G_1(q^2) \right. \\
&+  Q_{\mu}Q_{\nu} \left(p^{\prime}_{\alpha}p_{\beta}-p\cdot p^{\prime}\eta_{\alpha\beta}\right)G_2(q^2) \\
&+  \left.  \left(q_{\mu}q_{\nu} - q^2 \eta_{\mu\nu}\right)\left(p^{\prime}_{\alpha}p_{\beta}-p\cdot p^{\prime}\eta_{\alpha\beta}\right)G_3(q^2) \right]
\label{matrixelement}
\end{align}
where
\begin{align}
\nonumber
\mathcal{M}^0_{\mu\nu,\alpha\beta}&=p_{\mu}^{\prime} p_{\nu} \eta_{\alpha\beta} + p_{\mu} p^{\prime}_{\nu} \eta_{\alpha\beta} + \eta_{\mu\nu} p^{\prime}_{\alpha} p_{\beta} - p_{\mu} p^{\prime}_{\alpha} \eta_{\nu\beta} - p_{\mu}^{\prime} p_{\beta} \eta_{\alpha\nu} - p_{\nu} p^{\prime}_{\alpha} \eta_{\mu\beta} \\
&- p_{\nu}^{\prime} p_{\beta} \eta_{\alpha\mu} + p \cdot p^{\prime} (\eta_{\mu\alpha} \eta_{\beta\nu} + \eta_{\mu\beta} \eta_{\nu\alpha} - \eta_{\mu\nu} \eta_{\alpha\beta})
\end{align}
is the tree level matrix element and $G_{1,2,3}$ are form-factors.

We can extract this result from the energy momentum tensor found in the previous section. Unlike the effective action, the photons are dynamical in the matrix element compuation and thus we include the field-strength renormalization graphs shown in Fig. [\ref{Leg}]. These remove the dependence on the unphysical parameter $\mu$ and bring in mass singularities, and we have evaluated using the off-shellness condition $p^2=p'^2=\lambda^2$ to regulate these. The net effect is to replace the $\mu^2$ dependence within the logarithm with $\lambda^2$. The results for the massless
conformally coupled scalar are
\begin{align}
G_1 = 1 + e^2 b_s \ln(q^2/\lambda^2), \quad G_2 =  \frac{e^2}{96\pi^2 q^2}, \quad G_3 = -\frac{e^2}{96\pi^2 q^2} \ \ .
\label{scalar}
\end{align}
Note also the pole, $1/q^2$, in $G_2,~G_3$, which we also saw in the effective action. The equivalent result for a massless fermion \cite{Berends} corresponds to
\begin{align}
G_1 = 1 + e^2 b_f\ln(q^2/\lambda^2), \quad G_2 =  -\frac{e^2}{48\pi^2 q^2}, \quad G_3 = -\frac{e^2}{48\pi^2 q^2} \ \ .
\label{fermion}
\end{align}

We note that the trace anomaly relation emerges correctly in both cases, in that
\begin{align}
\langle \gamma(p')|T^{\mu}_{~ \mu}|\gamma(p)\rangle = \epsilon_\beta^*(p')\epsilon_\alpha(p)\left[   \left(p^{\prime}_{\alpha}p_{\beta}-p\cdot p^{\prime}\eta_{\alpha\beta}\right)q^2 \left(- G_2(q^2) - 3 G_3(q^2)\right) \right]
\label{tracematrixelement}
\end{align}
with
\begin{align}
q^2 \left(-G_2(q^2)- 3 G_3(q^2)\right) = \frac{\beta^{(s,f)}}{e} \ \ .
\label{tracenumbers}
\end{align}
In each case, the result is consistent with the relation
\begin{align}
T_\mu^{~ \mu} = \frac{\beta^{(s,f)}}{2e}F_{\mu\nu}F^{\mu\nu}
\end{align}
with the appropriate $\beta$ function. Although the matrix element has a $1/q^2$ pole, the trace is a constant.

\begin{figure*}
\centering
\begin{tabular}{cc}
\epsfig {file=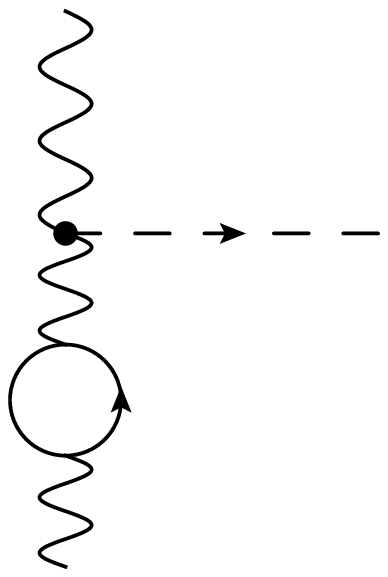,width=0.15\linewidth,clip=}& \quad\quad\quad
\epsfig {file=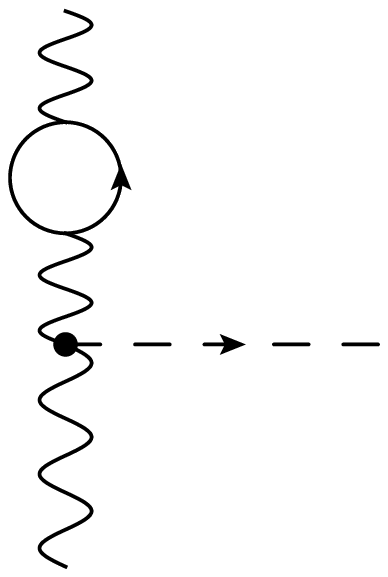,width=0.15\linewidth,clip=}
\end{tabular}
\label{Leg}
\caption{Photon self-energy diagrams needed for the matrix element.}
\end{figure*}

\section{Gravity and a non-linear completion of the action}

The connection between the non-local effective action and the trace anomaly is more obvious if we construct a non-linear form of the action using gravitational curvatures. There has been a lot of controversy in the literature about the correct form of the non-local action that gives rise to the anomaly. Some authors, see for example \cite{Mazur, Bonora, Giannotti}, argue for the Riegert action first obtained in \cite{Riegert,Fradkin} while others dismissed it based on several arguments \cite{Deser2, DeserS, Erdmenger} and proposed alternative forms. Moreover, another group of authors has used a renormalization group approach to argue that both forms exist in the effective action \cite{Codello}. One might try developing a non-linear completion based on the perturbative result \cite{Basem}, however this opens up extra questions about general covariance and uniqueness of the result. The answer to these questions will be addressed collectively in a companion publication \cite{Donoghue}.

When dealing with massive charged fields, the covariant form involving the curvatures could readily be found by one of two ways; non-linear completion or heat kernel methods. For massive fields, all Lagrangians are local and the expansion in the curvatures coincides with the energy or derivative expansion - higher powers of the curvature involve higher derivatives. To shed light on the difficulties of the construction when dealing with non-locality, we review a local action given by Drummond and Hathrell \cite{Drummond} corresponding to the one-loop effect of a massive charged fermion
\begin{align}\label{local}
\Gamma_{local}[g,A] = \frac{e^2}{m^2} \int d^4x \, \sqrt{g} \, \bigg[l_1 \, F_{\mu\nu}F^{\mu\nu} R + l_2 \, F_{\mu\sigma}F_\nu^{~\sigma} R^{\mu\nu} + l_3 \, F^{\mu\nu} F^\alpha_\beta  R_{\mu\nu\alpha}^{~ ~ ~ ~ \beta} + l_4\, \nabla_{\mu} F^{\mu\nu} \nabla_{\alpha} F^{\alpha}_{\, \nu} \bigg]
\end{align}

These operators comprise a complete basis up to third order in the generalized curvature expansion. In \cite{Drummond} they were determined using the two methods mentioned above; matching the above operators onto the perturbative calculation of \cite{Berends} in the low-energy limit and using the Schwinger-DeWitt technique to compute the heat kernel. Indeed the outcome of the two methods agreed, with the result
\begin{align}
l_1 = - \frac{1}{576 \pi^2}, \quad l_2 = \frac{13}{1440 \pi^2}, \quad l_3 = -\frac{1}{1440 \pi^2}, \quad l_4 = - \frac{1}{120 \pi^2} \ \ .
\end{align}

With non-local actions the curvature expansion is not equivalent to the
derivative or energy expansion because the calculations require factors of $1/q^2$ or $1/\Box$. Higher powers of $(1/\Box)R$ are not suppressed in the energy
expansion. Since there is no mass scale in the problem, derivatives acting on curvatures can not be deemed small and thus all powers of derivatives must be taken into account. One can think of the non-local form as a non-analytic expansion summarizing the results of a one-loop calculation. Nevertheless, the curvature expansion as in Eq. (\ref{local}) is useful because it accommodates the general covariance of the theory in a more explicit fashion.

%The result can be obtained from
%the perturbative calculation by a matching procedure - ensuring that the curvature invariants reproduce the perturbative result when expanded to first order.
%We refer to this matching as the {\em non-linear completion} of the effective action. This is achieved by proposing all possible operators with appropriate dimension and symmetries. We write the resulting effective action in quasi-local form.

In the local expansion the term involving the constant $l_4$ in Eq. (\ref{local}) is the only term which survives in flat space. It comes from the vacuum polarization and is the analogue of the
$\ln \Box$ of our non-local form. However, this coefficient has no relation to the beta function. For the other terms, the factors of $1/m^2$ have to be
replaced by a different factor with the same dimensionality. This can be done schematically by replacing $1/m^2$ by $1/\Box$ in Eq. (\ref{local}). The $1/m^2$ is the leading term in the low-energy expansion of a massive propagator, and thus for massless particles $1/\Box$ is the obvious generalization. Of course, the replacement is not exact, and we need to adjust the coefficients to match the perturbative result.

We find the following form to be the most informative
\begin{align}
\Gamma_{anom.}[g,A] = \int d^4x \, \sqrt{g} \left[n_R F_{\rho\sigma}F^{\rho\sigma}\frac{1}{\Box} R  + n_C F^{\rho\sigma} F^{\gamma}_{\, \, \lambda} \frac{1}{\Box}C_{\rho\sigma\gamma}^{\quad \lambda} \right]  \ \ .
\label{nonlinearcompletion}
\end{align}
In this basis, $\Box = g^{\mu\nu}\, \nabla_\mu \nabla_\nu$ is the covariant d' Alembertian and $C_{\rho\sigma\gamma}^{\quad \lambda} $ is the Weyl tensor which in 4$D$ reads
\begin{align}\label{weyl}
C_{\mu\nu\alpha\beta} = R_{\mu\nu\alpha\beta} - \frac12 \big(g_{\mu\alpha} R_{\nu\beta} - g_{\mu\beta} R_{\nu\alpha} - g_{\nu\alpha} R_{\mu\beta} + g_{\nu\beta} R_{\mu\alpha} \big) + \frac{R}{6} \big(g_{\mu\alpha} g_{\nu\beta} - g_{\mu\beta} g_{\nu\alpha}\big)
\end{align}
and
\begin{align}
n_R^{(s,f)} = - \frac{\beta^{(s,f)}}{12 e}, \quad n_C^s =  - \frac{e^2}{96 \pi^2}, \quad n^f_C = \frac{e^2}{48 \pi^2} \ \ .
\end{align}
The term with the Weyl tensor is unrelated to the beta function and the trace anomaly. The term involving the scalar curvature in the form $(1/\Box)R$ is the
nonlinear completion of the $1/\Box$ effects which leads to the conformal anomaly above. The latter is consistent with the leading part of the Riegert action  whose non-local piece reads
\begin{align}
\Gamma_{Riegert} = \frac{b}{4} \int d^4x \, \sqrt{g} \, F^2  \frac{1}{\Delta_4} \left(E - \frac{2}{3}\Box R \right)
\end{align}
where $E$ is the 4$D$ Gauss-Bonnet topological invariant and $\Delta_4$ is the fourth order operator \cite{Riegert}
\begin{align}
\Delta_4 = \Box^2 - 2 R^{\mu\nu}\nabla_\mu \nabla_\nu + \frac23 R \Box^2 - \frac13 (\nabla^\mu R)\nabla_\mu \ \ .
\end{align}
The Riegert action has additional contributions which are purely gravitational that we do not display. One immediately sees that the piece relevant for a linear expansion around flat space has the required form $F^2(1/\Box)R$ with $b = \beta/2e$. This aspect of the effective action was noticed before in \cite{Giannotti} as well.

\section{Quantum equivalence principle violation}

Quantum loops will upset the predictions of classical general relativity.
In this section, we display the quantum corrected formula for the bending angle of light and show the violation of the equivalence principle. The classical prediction of general relativity can be found in almost every textbook on general relativity. There is no reliable fully quantum treatment that can be applied
to the bending of light. We follow the semiclassical approach presented in \cite{Bohr}. The inverse Fourier transform of the amplitude is first obtained, from which one can define a semiclassical potential describing the interaction between a photon and a massive object like a star. This allows the bending angle to be computed via
\begin{align}\label{bendangle}
\theta \approx \frac{b}{E} \int_{\infty}^{\infty} du \, \frac{V^\prime(b\sqrt{1+u^2})}{\sqrt{1+u^2}}
\end{align}
where $b$ is the classical impact parameter and $E$ is the photon energy. Although this formula might look naive, it was shown in \cite{Bohr} that it indeed yields the correct result for the post-Newtonian correction to the bending angle when gravitaton loops are considered.

Because there are no completely massless charged particles\footnote{However, note that in the early universe above the electroweak phase transition, the elementary particles are massless.}, our result would only apply in the real world at energies far above the particle mass. However, it is interesting as a theoretical laboratory. What aspects of the equivalence principle can be violated by quantum effects? As a technical aspect, we allow the mass to provide an infrared cutoff to the infrared singularity of the energy-momentum matrix element. The coupling of photons to gravity is given by the one-loop energy-momentum tensor given in the previous section with $\lambda$ replaced by $m$.
\begin{figure*}
\centering
\begin{tabular}{cc}
\epsfig{file=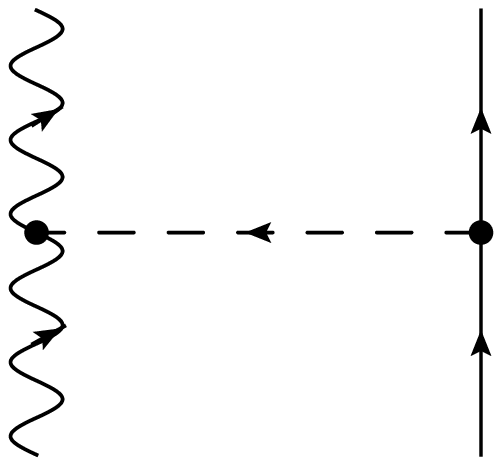,width=0.2\linewidth}\quad \quad \quad \quad \quad
\epsfig{file=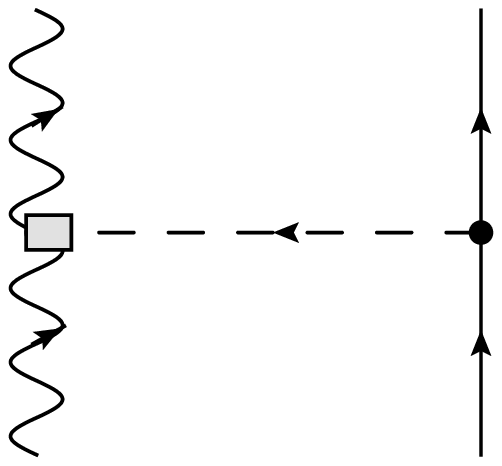,width=0.2\linewidth}
\end{tabular}
\caption{Gravitational scattering of a photon off a static massive target. The diagram on the left is the tree level process, while the square in the right diagram represents the non-local effects.}
\label{LO}
\end{figure*}

Since we work in the static limit, the scalar particle mass is large compared to the momentum transfer $M_{\odot} \gg |\bold{q}|$ and so we ignore the recoil. We also remind that the polarization vectors for physical photons are purely spatial and thus the amplitude takes the simple form
\begin{align}\label{amp}
\mathcal{M} = \frac{(\kappa M_{\odot} )^2}{2 \bold{q}^2} \left[1 - \frac{\beta^{(s,f)}}{e}\ln\left(\frac{\bold{q}^2}{\mu^2}\right) \right] \left (E^2 \bold{\epsilon}^{\star} \cdot \bold{\epsilon}(1 + \cos \theta) - \bold{k} \cdot \bold{\epsilon}^{\star} \bold{k}^{\prime} \cdot \bold{\epsilon} \right)
\end{align}
where $E$ is the photon energy, $\bold{k}$ is the incoming 3-momentum, $\bold{k}^\prime$ is the outgoing 3-momentum and the polarization vectors are purely spatial.

It is convenient to work with circularly polarized photons, and we find that the helicity conserving amplitude includes the contribution of the logarithm, yielding
\begin{align}
\mathcal{M}(++) = \mathcal{M}(--) = \frac{(\kappa M_{\odot}E )^2}{2 \bold{q}^2} \left[1 - \frac{\beta^{(s,f)}}{e}\ln\left(\frac{\bold{q}^2}{m^2}\right) \right] (1+\cos \theta)
\end{align}
In the non-relativistic limit, the semiclassical potential is simply
\begin{align}
V(r) = -\frac{1}{4M_{\odot}E} \int \frac{d^3q}{(2\pi)^3} \, e^{i \bold{q}\cdot \bold{x}} \mathcal{M}(q)
\end{align}
where the prefactor accounts for non-relativistic normalization. Employing the following relations,
\begin{align}
\int \frac{d^3\vec{q}}{(2\pi)^3} \frac{e^{-i \bold{q} \cdot \bold{r}}}{\bold{q}^2} \ln\left(\frac{\bold{q}^2}{m^2}\right) = -\frac{\ln(m r) +\gamma_E}{2\pi r}, \quad \int \frac{d^3\vec{q}}{(2\pi)^3} e^{-i \bold{q} \cdot \bold{r}} \ln\left(\frac{\bold{q}^2}{m^2}\right) = -\frac{1}{2\pi r^3}, \quad \cos \theta = 1-\frac{\bold{q}^2}{2E^2}
\end{align}
we simply find
\begin{align}
V_{++}(r) = V_{--}(r) = -\frac{2G M_{\odot}E}{r} + \frac{16 \pi G M_{\odot}}{E} \delta^{(3)}(\bold{x}) + \frac{4\beta G M_{\odot} E}{er} \left(\frac{1}{4E^2 r^2} - \ln mr - \gamma_E \right)
\end{align}
Notice in particular that the corrections to the Newtonian piece are not necessarily attractive. The short-range delta function does not lead to any modifications to the bending angle. Using Eq. (\ref{bendangle}), we find
\begin{align}
\theta_{non-flip} \approx \frac{4GM_\odot }{b} + \frac{8 \beta G M_\odot}{eb} \left(\ln mb + \gamma_E - \ln 2\right) - \frac{4 \beta G M_\odot}{e E^2b^3}
\end{align}

In contrast to this, the $1/q^2$ portion of the energy momentum tensor leads to helicity flip amplitudes. Here, one finds the result
\begin{align}
\mathcal{M}(+-) &= \mathcal{M}(-+) = - \frac{(\kappa e M_{\odot} E)^2}{\bold{q}^2} b_s + \frac{(\kappa e M_{\odot})^2}{4} b_s
\end{align}
for bosons and
\begin{align}
\mathcal{M}(+-) &= \mathcal{M}(-+) = \frac{(\kappa e M_{\odot} E)^2}{\bold{q}^2} b_f + \frac{(\kappa e M_{\odot})^2}{4} b_f
\end{align}
for fermions. This result has interesting features; first of all the sign in front of the Coulomb-like piece is different for both species. Moreover, the $1/q^2$ terms do not modify the helicity non-flip part of the amplitude. Thus the non-relativistic potential is spin-dependent. If we proceed with the calculation of the
bending angle, we find
\begin{align}
\theta_{flip} \approx \begin{cases}
- 4e^2 b_s G M_{\odot}/b, \quad \textrm{bosons} \\
4e^2 b_f G M_{\odot}/b, \quad \textrm{fermions}\\
\end{cases}
\end{align}
The interpretation of this result is less clear. However, the overall picture is clear: quantum physics has modified the classical prediction for light bending. In particular, photons of different energies will follow different trajectories.

\section{Conclusion}

We have been discussing low energy aspects of the conformal (trace) anomaly of QED using the one-loop effective action obtained by integrating out the massless charged particles. This is non-local because of the long distance propagation of the massless particles. However, after renormalization it is this non-local object that encodes the information on the anomaly. We also constructed the non-local energy-momentum tensor quadratic in the gauge field. This has the correct non-vanishing trace arising from a $1/q^2$ pole, which nevertheless yields a local trace. In the effective action, both the $\log \Box$ and $1/\Box$ terms were required, with the log piece being related to scale symmetry and the  $1/\Box$ piece being related to conformal symmetry. These non-local terms are interesting in their own right. For example, we showed that such corrections lead to an energy dependence of the bending of light, signaling a violation of some classical versions of the Equivalence Principle.

Another aspect of our exploration is an initial construction of the non-local action for a curved background, the correct form of which has been an ongoing controversy since the seminal work on gravitational anomalies by Deser, Isham and Duff \cite{DDI}. This construction constitutes a fundamental ingredient if one wants to consider the effects of the anomaly on various gravitational phenomena beyond the linear approximation. Over the years, multiple authors have investigated the effects of anomalies on different phenomena ranging from cosmology and astrophysics \cite{Labun,Thomas,Mottola,Bilic,Schutzhold,Hawking} to black holes \cite{Fursaev,Aros}. We will continue the discussion of the covariant form of the effective action in \cite{Donoghue}.

\section{\bf Acknowledgments}
We would like to thank A. Codello, S. Deser, E. Mottola and A. O. Barvinsky for useful discussions. This work has been supported in part by the U.S. National Science Foundation Grant No. PHY-1205896.

\appendix

\section{Scale currents}

 Let us give a quick review of scale and conformal symmetries in a bit more detail than described in the introduction. In general the consequence of dilatation symmetry is to generate a current
\begin{align}
J_{Noether}^{\mu} = \Theta^{\mu}_{\, \nu} x^{\nu} - j^\mu
\end{align}
where $j^\mu$ is called the virial current and $\Theta_{\mu\nu}$ is the {\em canonical} energy-momentum tensor. Scale symmetry then implies that
\begin{align}
\partial_\mu J_{Noether}^{\mu} = \Theta^{\mu}_{\, \mu}  - \partial_\mu j^\mu
\end{align}
For example, if we apply Noether's theorem to $S_{EM}$ we find
\begin{align}
J_{Noether}^{\mu} = \Theta^{\mu}_{\, \nu} x^{\nu} - F^{\mu\alpha}A_{\alpha}
\end{align}
where $\Theta_{\mu\nu}$ is
\begin{align}
\Theta_{\mu\nu} = \frac{1}{4} g_{\mu\nu} F_{\alpha\beta}F^{\alpha\beta} - F_{\mu\alpha}\partial_{\nu} A^{\alpha} \ \ .
\end{align}
The current is easily seen to be conserved upon using the classical equation of motion, but notice that it looks quite different from the dilatation current in Eq. (\ref{divD}). Moreover, the {\em asymmetric} canonical energy-momentum tensor is not the same as $T_{\mu\nu}$ quoted in the same equation. The trick is to use scale invariance to construct an {\em improved} traceless tensor much like using the Belinfante procedure for finding a symmetric energy-momentum tensor exploiting the Lorentz invariance of the theory. These aspects are well explained in \cite{DiFrancesco,Weinberg}. The procedure is to judiciously add a conserved symmetric second-rank tensor to form the Belinfante tensor such that its trace reads
\begin{align}
T_{\mu}^{~\mu} = \partial_{\mu} J_{Noether}^{\mu}\big|_{off-shell}
\end{align}
and hence the {\em improved} tensor $T_{\mu\nu}$ will be traceless on-shell. For electromagnetism, the Belinfante procedure yields the desired tensor without any further modifications\footnote{Note that the energy-momentum tensor is traceless even off-shell.}
\begin{align}\label{classicaleom2}
T_{\mu\nu} = - F_{\mu\sigma} F_{\nu}^{\sigma} + \frac{1}{4} g_{\mu\nu} F_{\alpha\beta}F^{\alpha\beta} \ \ .
\end{align}
With this object in hand, Eq. (\ref{divD}) defines the dilatation current. When coupled to gravity, the photon action is conformally invariant.

A similar story holds for the scalar field, starting from the Lagrangian of Eq. (\ref{scalaraction}). For the minimally coupled field, the energy momentum tensor is not traceless and the dilatation current is
\begin{align}
J_{Noether}^{\mu} = T^{(\xi=0)~\mu}_{\, \nu} x^{\nu} - [\phi^\star \partial^\mu \phi + (\partial^\mu \phi^\star)\phi]
\end{align}
However, if we use the improved energy momentum tensor with conformal coupling, the energy momentum tensor is now traceless

\begin{align}
T^{(\xi=1/6)~\mu}_{\, \mu} =0
\end{align}
and we do not need the virial current. The scalar field is only conformally invariant for $\xi=1/6$.

\section{Reduction of the triangle and bubble integrals}\label{reduction}
\subsection{\bf Bubbles}
\begin{align}
\int \frac{d^Dk}{(2\pi)^D} \frac{k^\mu}{(k^2+i0)((k+l)^2+i0)} &= -\frac{1}{2} l^\mu I_2(l) \\
\int \frac{d^Dk}{(2\pi)^D} \frac{k^\mu k^\nu}{(k^2+i0)((k+l)^2+i0)} &= \frac{1}{4(D-1)} \left[D l^\mu l^\nu - l^2 \eta^{\mu\nu}\right] I_2(l) \\
\int \frac{d^Dk}{(2\pi)^D} \frac{k^\mu k^\nu k^\alpha}{(k^2+i0)((k+l)^2+i0)} &= \frac{1}{8(D-1)}\left[l^2 (\eta^{\mu\nu} l^\alpha + \eta^{\mu\alpha} l^\nu + \eta^{\alpha\nu} l^\mu ) - (D+2) l^\mu l^\nu l^\alpha \right]  I_2(l)
\end{align}
where $l$ is an arbitrary four-momentum and $I_2$ is the scalar bubble function
\begin{align}
I_2(p) = \int\frac{d^Dk}{(2\pi)^D} \frac{1}{(k^2+i0) ((k+p)^2 + i0)}
\end{align}
\subsection{\bf Triangles}
\begin{align}
\int \frac{d^Dk}{(2\pi)^D} \frac{k^\mu}{(k^2+i0)((k+l)^2+i0)((k+l^\prime)^2+i0)} &= A Q^\mu \\
\int \frac{d^Dk}{(2\pi)^D} \frac{k^\mu k^\nu}{(k^2+i0)((k+l)^2+i0)((k+l^\prime)^2+i0)} &= B \eta^{\mu\nu} + C Q^\mu Q^\nu + D q^\mu q^\nu \\\nonumber
\int \frac{d^Dk}{(2\pi)^D} \frac{k^\mu k^\nu k^\alpha}{(k^2+i0)((k+l)^2+i0)((k+l^\prime)^2+i0)} &= E (Q^\mu \eta^{\nu\alpha} + \text{perm}) + F Q^\mu Q^\nu Q^\alpha + G (Q^\mu q^\nu q^\alpha + \text{perm})\\\nonumber
\int \frac{d^Dk}{(2\pi)^D} \frac{k^\mu k^\nu k^\alpha k^\beta}{(k^2+i0)((k+l)^2+i0)(k+l^\prime)^2+i0)}&= H (\eta^{\mu\nu} \eta^{\alpha\beta} + \text{perm}) + I (\eta^{\mu\nu} Q^\alpha Q^\beta + \text{perm}) + J (\eta^{\mu\nu} q^\alpha q^\beta + \text{perm})\\
&+ K Q^\mu Q^\nu Q^\alpha Q^\beta + L q^\mu q^\nu q^\alpha q^\beta + M (Q^\mu Q^\nu q^\alpha q^\beta + \text{perm})
\end{align}
where
\begin{align}
l^2 = l^{\prime 2} = \lambda^2 \rightarrow  0, \quad Q = l+ l^\prime \quad q = l-l^\prime
\end{align}
We ignored any analytic dependence on $\lambda^2$, and only retained it inside logarithms. The different coefficients read
\begin{align}
\nonumber
A &= \frac{1}{q^2} (I_2(q) - I_2(l)),\quad B = \frac{1}{2(D-2)} I_2(q), \quad C = \frac{1}{q^2} \left(\frac{1}{4}I_2(l) - \frac{D-3}{2(D-2)} I_2(q)\right)\\\nonumber
D &= \frac{1}{q^2} \left(\frac{1}{4}I_2(l) - \frac{1}{2(D-2)} I_2(q)\right), \quad E = -\frac{1}{4(D-1)}I_2(q)\\\nonumber
F &= \frac{1}{4q^2(D-1)}\left((D-3) I_2(q) - \frac{D}{4} I_2(l) \right), \quad G = \frac{1}{4q^2(D-1)} \left(I_2(q) - \frac{D}{4} I_2(l)\right) \\\nonumber
H &= -\frac{q^2}{8D(D-1)}I_2(q), \quad I = \frac{1}{8D}I_2(q), \quad J = \frac{1}{8D(D-1)}I_2(q) \\\nonumber
K &= \frac{1}{8q^2}\left(\frac{D+2}{8(D-1)}I_2(l) - \frac{D-3}{D}I_2(q)\right), \quad L = \frac{1}{8q^2(D-1)}\left(\frac{D+2}{8}I_2(l) - \frac{3}{D}I_2(q)\right)\\
M &= \frac{1}{8q^2}\left(\frac{D+2}{8(D-1)}I_2(l) - \frac{1}{D}I_2(q)\right)
\end{align}

\end{document}